# Comparative Analysis of Data-Driven Predictive Control Strategies

Sohrab Rezaei
*Department of Electrical Engineering*
*K. N. Toosi University of Technology*

Ali Khaki-Sedigh
*Department of Electrical Engineering*
*K. N. Toosi University of Technology*
Shariati Ave. Seyed Khandan Bridge,
Tehran, Iran, 16317-14191
sedigh@kntu.ac.ir

***Abstract*—This paper compares data-driven predictive control strategies by examining their theoretical foundations, assumptions, and applications. The three most widely recognized and consequential methods, Data Enabled Predictive Control, Willems-Koopman Predictive Control, Model-Free Adaptive Predictive Control are employed. Each of these strategies is systematically reviewed, and the primary theories supporting it are outlined. Following analysis, a discussion is provided regarding their fundamental assumptions, emphasizing their influence on control effectiveness. A numerical example is presented as a benchmark for comparison to enable a rigorous performance evaluation.**

## I. Introduction

Data-driven control is an emerging paradigm in control systems design. The predictive nature of some of the data-driven control methodologies make them appropriate for practical applications. However, a comparative study of such methodologies will enhance the designers' ability to select the appropriate method for their particular application. This paper studies and compares the data-driven predictive controllers from a closed-loop performance perspective.

Since the late 1970s, much interest has been shown in predictive control for industrial applications [1]. Model Predictive Control (MPC) has been a widely applied control method [2]. It is capable of optimizing a cost function in the presence of constraints. The model-based approaches, which meet high control performance and robustness to uncertainties and disturbances, directly depend on the model's accuracy, impacting the efficiency and stability of model-based predictive approaches. Model-based methods, which rely on first principles and system identification, are showing signs of incompetence in handling control challenges in complex systems. This trend has led to development and establishment of the Data-Driven Control (DDC) theory, addressing the need for effective control solutions, notably when accurate process models are lacking [3]. These methods can handle the situation with minimal assumptions by only relying on input-output data without the need for structural information. Data-driven methods aim to enhance the efficiency of model-based approaches and rely on data to bypass the need for an accurate model. Several data-driven and data-based predictive control strategies have been proposed, as evidenced by notable works such as [4] and [5]. However, for the sake of conciseness and to streamline the paper's focus, this study undertakes a comprehensive analysis and comparison of three prominent predictive control methodologies: Data Enabled Predictive Control (DeePC), also known as Data-Driven Model Predictive Control, Willems-Koopman Predictive Control (WKPC), and Model-Free Adaptive Predictive Control (MFAPC). It is essential to note that while the terms "data-driven" and "data-based" may seem interchangeable, their distinctions are defined in detail in [6]. Based on Willems' Fundamental Lemma [7], it is possible to characterize the behavior of linear time-invariant (LTI) systems using a persistently exciting input-output data set. Willems' fundamental lemma for unknown linear systems forms the basis of the DeePC algorithm. By employing the behavioral model instead of the parametric model in constraints, this method utilizes the traditional objective function of the MPC. [8] and [9] extended this method to nonlinear systems.

Another method involves combining the Koopman operator with Willems' fundamental lemma. In 1931, Koopman introduced his operator, which can map nonlinear systems to infinite-dimensional linear systems [6]. Due to the difficulty of locating Koopman operators and the infinite dimension of the lifted linear system, the Koopman operator first lost interest in the control community. However, in [10] and [11], the potential effectiveness of the Koopman operator has been suggested. Nevertheless, computation of the Koopman operator has become straightforward with the development of the efficient numerical method of Dynamic Mode Decomposition (DMD) and Extended Dynamic Mode Decomposition (EDMD). Therefore, this method is suited to the control community's requirements. An overview of WKPC is given in [6].

This is the accepted version of the paper:
S. Rezaei, A. Khaki-Sedigh,
"Comparative Analysis of Data-Driven Predictive Control Strategies"
, 2023 9th International Conference on Control, Instrumentation and Automation (ICCIA), Tehran, Iran,
Dec. 20–21, 2023. DOI: 10.1109/ICCIA61416.2023.10506351, Published version available at: https://ieeexplore.ieee.org/document/10506351

The model-free adaptive control (MFAC) technique is one of the most popular data-driven control algorithms [6]. MFAC utilizes a pseudo model driven from the system's input-output data to provide a control signal by minimizing a specified cost function. The pseudo model is a virtual equivalent dynamic linearized data model built at each sampling time. Model-Free Adaptive Predictive Control (MFAPC) uses the pseudo model to predict the system's behavior along the prediction horizon and has the advantages of MPC and MFAC. It is possible to apply this methodology to a class of unknown nonaffine nonlinear systems [2]. In [2], three types of pseudo models are introduced that are Compact Form Dynamic Linearization (CFDL), Partial Form Dynamic Linearization (PFDL), and Full Form Dynamic Linearization (FFDL). These three pseudo models are also used to develop MFAPC.

## II. THEORETICAL FOUNDATIONS

In this section, relevant theories are explored and analyzed in depth, including an examination of their foundational principles, key concepts, and application to the context of this study.

### A. Willems' Fundamental Lemma

The Hankel matrix of depth $L$ for a time series data $\{w_k\}_{k=1}^{T} \in \mathcal{B}$ ($\mathcal{B}$ is assumed to be a linear time-invariant unknown system), is defined as follows:

$$H_L(w) = \begin{bmatrix} w_1 & w_2 & \cdots & w_{T-L+1} \\ w_2 & w_3 & \cdots & w_{T-L+2} \\ \vdots & \vdots & \ddots & \vdots \\ w_L & w_{L+1} & \cdots & w_T \end{bmatrix} \quad (1)$$

Time series data (w) is persistently exciting of order $L$ if $H_L(w)$ is full row rank (assuming $T - L \geq L - 1$).

A behavioral model can be represented under the following assumptions.

**Assumption 1:** System $\mathcal{B}$ is controllable in the sense of behavioral system theory [12].

**Assumption 2:** The input component of $w$ is persistently exciting of order $l + n(\mathcal{B})$. Where $n(\mathcal{B})$ is an upper bound of the system's order, $l$ is the system's lag.

The fundamental lemma [7] shows that with the sequence of $\{u_k^d, y_k^d\}_{k=1,\dots,T}$, any sequence of the system $\{u_k, y_k\}_{k=1,\dots,L}$ could be mapped with a vector $g \in \mathbb{R}^{T-L+1}$ as follows, and it is a behavioral system model that is driven out from system input-out data:

$$\begin{bmatrix} H_L(u^d) \\ H_L(y^d) \end{bmatrix} g = \begin{bmatrix} u \\ y \end{bmatrix} \quad (2)$$

### B. Koopman Operator

A discrete-time autonomous nonlinear system,

$$x_{k+1} = f(x_k) \quad (3)$$

is considered, where $f$ models the nonlinear dynamics. The Koopman Operator $K$ is an infinite-dimensional linear operator [13], that acts on nonlinear functions as:

$$K\psi := \psi \circ f \quad (4)$$

where $\circ$ is the composition operator, and $\psi: \mathbb{R}^{n_x} \to \mathbb{R}$ is called an observable. For the discrete-time nonlinear system (3) equation (4) becomes:

$$K\psi(x_k) = \psi f(x_k) = \psi(x_{k+1}) \quad (5)$$

Since we are dealing with a linear operator with infinite dimensions, we do not intend to use all observable functions. Instead, we can use only key observable functions that behave linearly. Using the Koopman operator's eigenfunctions, a set of distinct observables that behave linearly in time can be derived. A discrete-time Koopman eigenfunction $\phi$ corresponding to eigenvalue $\lambda$ satisfies:

$$\phi(x_{k+1}) = \phi(f(x_k)) = (K\phi)(x_k) = \lambda\phi(x_k) \quad (6)$$

With a collection of eigenfunctions, any observable lying within the span of these eigenfunctions can be decomposed into $\psi = \sum_i c_i(\psi)\phi_i$, where $c_k(\psi)$ is called the Koopman modes of $\psi$. Accordingly

$$K\psi = \sum_i c_i(\psi)\lambda_i\phi_i \quad (7)$$

with $\lambda_i$ denoting the eigenvalue of $\psi_i$. The Koopman operator provides a global linear representation; however, in the case of data-driven models, it is not possible to determine the Koopman eigenfunctions. Accordingly, DMD and EDMD approximate the eigenfunctions [6]. An EDMD algorithm approximating Koopman eigenfunctions using radial basis functions is proposed in [6]. In this way, it is possible to have a linear representation of the nonlinear system (3) using only input and output information.

### C. Data-Enabled Predictive Control

By combining the traditional MPC [1] and fundamental lemma (II.A), the DeePC algorithm for linear time-invariant systems is introduced in [8]. Furthermore, a slack variable $\sigma_y$ was introduced to the optimization problem to account for the linearization error of the behavioral model and noisy data. Hence, the norm of the $g$ is involved in defining the cost function to make it bounded. Additionally, in [9], an artificial set point $[u^s, y^s]$ is defined that is optimized online to perform tracking characteristics of the algorithm. The modified predictive control problem can be shown as follows:

$$\begin{aligned} \underset{g,u,y,\sigma_y,u^s,y^s}{minimize} \sum_{k=1}^{N} (&\|y_k - y^s(t)\|_Q^2 \\ &+ \|r_{t+k} - y^s(t)\|_S^2 \\ &+ \|u_k - u^s(t)\|_R^2) + \lambda_g\|g\|_2^2 \\ &+ \lambda_\sigma\|\sigma_y\|_2^2 \end{aligned} \quad (8)$$

$$subject\ to\ \begin{pmatrix} U_p \\ Y_p \\ U_f \\ Y_f \end{pmatrix} g = \begin{pmatrix} u_{ini} \\ y_{ini} \\ u \\ y \end{pmatrix} + \begin{pmatrix} 0 \\ \sigma_y \\ 0 \\ 0 \end{pmatrix}$$

$$\begin{bmatrix} u_{[N-Tini+1,N]} \\ y_{[N-Tini+1,N]} \end{bmatrix} = \begin{bmatrix} u^s_{Tini}(t) \\ y^s_{Tini}(t) \end{bmatrix}$$

$$u_k \in \boldsymbol{U},\ \forall k \in [1,N], y_k \in \boldsymbol{Y},\ \forall k \in [1,N]$$

where $N \in \mathbb{Z}_{>0}$ is prediction time horizon, $r$ is reference trajectory, $[u_{ini}, y_{ini}]$ past system's input-output data, input constraints $\boldsymbol{U} \subseteq \mathbb{R}^m$, output constraints $\boldsymbol{Y} \subseteq \mathbb{R}^l$, output cost matrices $Q \in \mathbb{R}^{l\times l}, S \in \mathbb{R}^{l\times l}$, and control cost matrix $R \in \mathbb{R}^{m\times m}$, and $u^s_{T_{ini}} = [u^s, \dots, u^s]_{m\times T_{ini}}$, $y^s_{T_{ini}} = [y^s, \dots, y^s]_{l\times T_{ini}}$ and $\lambda_g, \lambda_{\sigma_y} > 0$.

Closed-loop stability guarantees for linear time-invariant systems are available in [14], and for nonlinear systems are available in [9].

### D. Data-Driven Willems-Koopman Predictive Control

An unknown nonlinear system is considered as below:

$$\begin{aligned} x_{k+1} &= f(x_k, u_k) \\ y_k &= h(x_k) \end{aligned} \tag{9}$$

**Assumption 3:** The input signal $u$ in the system (9) is persistently exciting of order $l + n_p$ where $n_p$ is the number of Koopman eigenfunctions selected by the designer, and $l$ is system lifted system's lag.

**Assumption 4:** The lifted system using the Koopman operator is controllable in the sense of behavioral system theory.

Based on the fundamental lemma, and under assumptions 3 and 4 if there exists a sequence $\{z_k, u_k, y_k\}_{k=1,\dots,T}$ that $u$ is persistently exiting of order $l + n_p$ and $z_k = \psi(x_k)$ ($\psi: \mathbb{R}^{n_x} \to \mathbb{R}^{n_p}$ is Koopman eigenfunctions), then any sequence of the system $\{z_k, u_k, y_k\}_{k=1,\dots,L}$ could be mapped with a vector $g \in \mathbb{R}^{T-L+1}$ as follows:

$$\begin{pmatrix} Z \\ U \\ Y \end{pmatrix} g = \begin{pmatrix} z_{[t-Tini+1,t+N]}(t) \\ u_{ini} \\ u \\ y_{ini} \\ y \end{pmatrix} \tag{10}$$

where $Z, U, Y$ are the Hankel matrices of lifted states, inputs, and outputs, respectively. Considering the above results, the data-driven Willems-Koopman Predictive Control problem is defined as follows:

$$\begin{aligned} \underset{g,u,y}{minimize} \sum_{k=1}^{N} \big( &\|y_k - r_{t+k}\|_Q^2 \\ &+ \|u_k - u^s(t)\|_R^2 \big) \\ &+ \lambda_g \|g\|_2^2 \end{aligned} \tag{11}$$

$$subject\ to\ Eq. (10)$$

$$z_k(t) = \psi\big(x_k(t)\big)$$

$$u_k \in \boldsymbol{U},\ \forall k \in [1,N], y_k \in \boldsymbol{Y},\ \forall k \in [1,N]$$

where $\psi_i, i = 1, \dots, n_p$ are Koopman operator eigenfunctions. The designer selects $n_p$, the order of lifted space; other variables and penalty terms are the same as in the previous subsection. This paper uses the following radial basis function to utilize the approximation of Koopman eigenfunctions.

$$\psi_i(x_k) = \|x_k - c_i\|_2^2 \times log_{10}(\|x_k - c_i\|_2) \tag{12}$$

where $c_i$ is randomly selected for each iteration. Closed-loop stability analysis for this method is utilized in [15].

### E. Model-Free Adaptive Predictive Control

Consider a class of discrete-time nonlinear systems described by

$$y_{k+1} = f(y_k, \dots, y_{k-n_y}, u_k, \dots, u_{k-n_u}) \tag{13}$$

It is shown in [2] nonlinear system (13) can be transformed into a CFDL, PFDL, or FFDL data model, which can capture the input-output behavior of the true plant. Based on this data model, $N$-step-ahead prediction could be available using the estimation of the data model. In this paper, equations of CFDL-based predictive control are given, and due to the similarity of equations and lack of space, other methods are not mentioned, but they are available in [2].

The existence of the dynamic linearization data model is proved in [2] for Single Input, Single Output (SISO), and multivariable systems. The following assumptions are necessary for the exitance of the CFDL data model.

For the system (13) it is assumed that:

**Assumption 5:** The partial derivative of $f(.)$ in (13) concerning the $n_y + 1$th variable is continuous for all $k$.

**Assumption 6:** System (13) satisfies the generalized *Lipschitz* condition, for all $k$, that is

$$|y_{k_1+1} - y_{k_2+1}| \le b|u_{k_1} - u_{k_2}|$$

for $u_{k_1} \ne u_{k_2}$ and any $k_1 \ne k_2$, and $k_1, k_2 \ne 0$, where $b$ is a positive constant.

Consider nonlinear system (13) satisfying assumptions 5 and 6. If $|\Delta u_k| \ne 0$, then there exists a time-varying parameter $\phi_k \in \mathbb{R}$, called Pseudo Partial Derivative, (PPD) such that system (13) can be transformed into the following CFDL data model:

$$\Delta y_{k+1} = \phi_k \Delta u_k \tag{14}$$

with bounded $\phi_k$ for any time $k$.

The point is that PPD is a time-varying unknown parameter difficult to obtain analytically, so it must be estimated from input-output data. As a result, a cost function is considered, and optimizing it could produce an estimation of the $\phi_k$.

$$J_\phi = |y_k - y_{k-1} - \phi_k \Delta u_{k-1}|^2 + \mu|\phi_k - \hat{\phi}_{k-1}|^2 \tag{15}$$

where $\mu > 0$. Using the concept idea of the projection algorithm and minimizing cost function (15) gives the following estimation algorithm:

$$\hat{\phi}_k = \hat{\phi}_{k-1} + \frac{\eta \Delta u_{k-1}}{\mu + \Delta u_{k-1}^2}[\Delta y_k - \hat{\phi}_{k-1}\Delta u_{k-1}] \tag{16}$$

This method's stability analysis has to consider the following assumptions.

**Assumption 7:** For a given bounded desired output signal $r_{k+1}$, there exists a bounded control input $u_k^*$ such that the system output driven by $u_k^*$ is equal to $r_{k+1}$. Where $r_{k+1}$ is the desired system output.

**Assumption 8:** The sign of PPD is assumed unchanged for all $k$ and $\Delta u_k \neq 0$ and $|\phi_k| > \varepsilon$ is satisfied, where $\varepsilon$ is a small positive constant.

To satisfy assumption 8, a resetting condition is defined:

$$\hat{\phi}_k = \hat{\phi}_1 \quad IF\ |\hat{\phi}_k| < \varepsilon\ or\ sign(\hat{\phi}_k) \neq sign(\hat{\phi}_1)\ or\ |\Delta u_k| \leq \varepsilon \tag{17}$$

where the step factor $\eta \in (0,2]$ is added to make the algorithm more general and more flexible also, $\hat{\phi}_k$ is an estimation of$\phi_k$. To predict $\hat{\phi}_k$ for $N$ samples forward, in [2], implied that the multilevel hierarchical forecasting method possesses the best predictive error; in this work, we used its results, which gives the following relations.

$$\hat{\phi}_{k+j} = \theta_1(k)\hat{\phi}_{k+j-1} + \theta_2(k)\hat{\phi}_{k+j-1} + \cdots + \theta_{n_m}(k)\hat{\phi}_{k+j-n_m}$$
$$j = 1,2,\ldots,N-1$$

$$\hat{\phi}_{k+j} = \hat{\phi}_1,\ IF\ |\hat{\phi}_{k+j}| < \varepsilon\ or\ sign(\hat{\phi}_{k+j}) \neq sign(\hat{\phi}_1)$$
$$j = 1,2,\ldots,N-1$$

where $n_m$ is the fixed model order that is usually set to $n_m \in [2,7]$. Defining $\Theta(k) = [\theta_1(k),\ldots,\theta_{n_m}(k)]^T$, it is determined by the following equation:

$$\Theta(k) = \Theta(k-1) + \frac{[\hat{\phi}_{k-1},\ldots,\hat{\phi}_{k-n_m}]^T}{\delta + \|[\hat{\phi}_{k-1},\ldots,\hat{\phi}_{k-n_m}]^T\|^2} \times [\hat{\phi}_k - [\hat{\phi}_{k-1},\ldots,\hat{\phi}_{k-n_m}]\Theta(k-1)] \tag{18}$$

Resetting condition for $\Theta(k)$ is

$$\Theta(k) = \Theta(1),\ IF\ \|\Theta(k)\| \geq M$$

where $M$ is a positive constant selected by the designer.

For the $N$-step-ahead prediction problem, the following equations are given:

$$\mathbf{Y}_N(k+1) = [y_{k+1},\ldots,y_{k+N}]^T$$
$$\Delta\mathbf{U}_N(k) = [\Delta u_k,\ldots,\Delta u_{k+N-1}]^T$$
$$\boldsymbol{E}_k = [1,\ldots,1]^T_{1\times Nl}$$

$$\boldsymbol{A}_k = \begin{bmatrix} \phi_k & 0 & 0 & \cdots & & 0 \\ \phi_k & \phi_{k+1} & 0 & \cdots & & 0 \\ \vdots & \vdots & \ddots & \vdots & & \vdots \\ \phi_k & \phi_{k+1} & \cdots & \phi_{k+Nu-1} & \cdots & 0 \\ \vdots & \vdots & \vdots & \vdots & \ddots & 0 \\ \phi_k & \phi_{k+1} & \cdots & \phi_{k+Nu-1} & \cdots & \phi_{k+N-1} \end{bmatrix}$$

$$\mathbf{Y}_N(k+1) = \boldsymbol{E}_k y_k + \boldsymbol{A}_k \Delta\mathbf{U}_N(k)$$

In this work, where $N_u$ is the control input horizon, as implied in [2] to improve transient and tracking response $N_u$, could be considered as $N_u = N$.

The cost function for achieving control law is considered as below.

$$J_u = [\mathbf{Y}_N^*(k+1) - \mathbf{Y}_N(k+1)]^T[\mathbf{Y}_N^*(k+1) - \mathbf{Y}_N(k+1)] + \lambda\Delta\mathbf{U}_N^T(k)\Delta\mathbf{U}_N(k) \tag{19}$$

Where $\lambda > 0$ is weighting factor and $\mathbf{Y}_N^*(k+1) = [r_{k+1},\ldots,r_{k+N}]^T$. The optimality condition leads us to the control law:

$$\Delta\mathbf{U}_N(k) = [\hat{\boldsymbol{A}}_k^T\hat{\boldsymbol{A}}_k + \lambda\boldsymbol{I}_{N\times N}]^{-1}\hat{\boldsymbol{A}}_k^T[\mathbf{Y}_N(k+1) - \boldsymbol{E}_k y_k] \tag{20}$$

Thus, the control input at the current time $k$ is obtained according to the receding horizon principle as follows:

$$u_k = u_{k-1} + \gamma\Delta\mathbf{U}_N(k) \tag{21}$$

where $\gamma = [1,0,\ldots,0]_{N\times 1}$.

If discrete-time nonlinear system (13), satisfying assumptions 5, 6, 7, and 8, is controlled by CFDL–MFAPC for a regulation problem, that is, $y^*(k+1) = r_k = const.$, then there exists a constant $\lambda_{min} > 0$ such that the following properties hold for any $\lambda > \lambda_{min}$:

- The tracking error of the system converges to zero.
- Input and output signals are bounded sequences.

Stability analysis for CFDL-MFAPC is available in [2], and FFDL-MFAPC is available in [16]. Furthermore, stability analysis is available for multivariable systems in [17].

## III. COMPARISON OF THEORIES AND ASSUMPTIONS

In the case of nonlinear systems, these three methods are based on linearizing behavior of the true plant and driving the control signal using identified behavior using merely input-output data and minimum system assumptions and do not use any system's mathematical model. However, there are some fundamental differences between them. DeePC fits a linear behavioral model to the nonlinear system and collects linearization error to the slack variable. WKPC maps systems' trajectories to linear space, and MFAPC uses an estimated pseudo-linearized model. Assumptions 1, 4, and 7 are the same as they are addressing controllability of the system in the sense of behavioral theory. Assumptions 2, 3 and condition $\Delta u(t) \neq 0$ in assumption 8, all are about persistent excitation (PE) of the input signal. However, the main difference between them is the PE order. MFAPC needs a smaller PE order, as it just has $L_y + L_u$ (design parameters) to estimate that it is equal to one for MFAPC-CFDL. WKPC the largest PE order between these methods ($n_p$ in assumption 3 is always greater than $n(\mathcal{B})$ in assumption 2).

In the field of control system design, assumption 5 is a typical constraint for general nonlinear systems [2]. Based on assumption 6, a ceiling is imposed on the rate at which the control input influences the system's output. Based on this assumption, only stable systems are compatible with MFAPC-CFDL and MFAPC-PFDL, but not with MFAPC-FFDL because the generalized *Lipschitz* condition for this method [2] includes systems output from both sides of the inequality. To deal with unstable systems using CFDL or PFDL approach, in [6], a simultaneous perturbation stochastic approximation controller is used to stabilize the system, and the MFAC method is employed for tracking response of the system, and also in [18] the system is first stabilized by using an i-PID controller, and then controlled by an MFAC controller to achieve a good tracking response.

**Remark 1:** In assumption 8, it is imperative that the PPD sign remains known and unchanged, which means that the control direction has to be known and unchanged in the true plant. However, it is essential to recognize that this assumption does not apply to DeePC or WKPC. A simulation was carried out to investigate how DeePC and WKPC respond to system control direction changes. These two methods exhibit performance challenges when the data used for constructing Hankel matrices exclude information from the system's previous configuration before the control direction changes and after $T$ sampling time, their effectiveness in managing the system control task improves as the Hankel matrices assimilate newly acquired data, emphasizing the crucial role of the historical context within the dataset.

**Remark 2:** Based on theoretical principles and empirical insights gained from simulation studies, it becomes evident that MFAPC is appropriate to address time-varying and nonlinear systems due to its adaptive nature. In contrast, the two other methods, DeePC and WKPC, necessitate the availability of a persistently exciting dataset derived from a time-invariant system. When dealing with time-varying systems and opting to utilize DeePC or WKPC methodologies, an implicit assumption must be made that the system's transition rate is sufficiently slow to ensure the dataset encompasses the evolving system behavior. In the context of nonlinear systems, it is required that the dataset contains samples from all operating points of the true plant.

## IV. Numerical Example

In this section, the data-driven strategies of section II, are applied to a nonlinear pendulum as a benchmark to examine their performance. The nonlinear equation of the pendulum is given as follows:

$$\begin{aligned} \dot{x}_1 &= x_2 \\ \dot{x}_2 &= -\frac{\boldsymbol{g}}{\boldsymbol{r}} sin(x_1) - \frac{k}{m\boldsymbol{r}} x_2 + \frac{\tau}{m\boldsymbol{r}} \end{aligned} \tag{22}$$

$x_1$ is the pendulum's angle to the vertical line and is considered the system output, and $x_2$ is the angular velocity. $m$ is the pendulum mass and is equal to $1[kg]$, $\boldsymbol{r}$ is the radius of the pendulum and is considered to be $0.2[m]$, $\boldsymbol{g}$ is the standard acceleration of gravity for Earth and is considered to be $9.81[\frac{m}{s^2}]$, also, $\tau$ is the applied torque and is considered as the control input. It is worth mentioning that here, the system's model is only used to generate input-output data. Sampling time is $0.1[s]$. Tunable parameters for DeePC and WKPC are mentioned in Table 1.

*Table 1. DeePC and WKPC parameters*

| | $\boldsymbol{T}$ | $\boldsymbol{T_{ini}}$ | $\boldsymbol{N}$ | $\boldsymbol{Q}$ | $\boldsymbol{S}$ | $\boldsymbol{R}$ | $\boldsymbol{\lambda_g}$ | $\boldsymbol{\lambda_\sigma}$ |
|---|---|---|---|---|---|---|---|---|
| DeePC | $20[s]$ | $0.3[s]$ | $0.5[s]$ | 100 | 300 | 10 | 50 | $10^7$ |
| WKPC | $20[s]$ | $0.2[s]$ | $0.5[s]$ | 1 | - | 0.1 | 0.1 | - |

For WKPC, $n_p$ is set to 10.

In our case study, MFAPC-CFDL exhibits more sensitivity to $\lambda$ values than others. With $\lambda = 0.7$, the controller behaved more smoothly but had a slower raise-time, while $\lambda = 0.25$ resulted in a faster raise-time but oscillating around the set point. Table 2 provides other parameters for the MFAPC-CFDL algorithm.

In this study, we performed simulations on a Linux-based operating system (Pop!_OS) using MATLAB 2023a. The hardware setup featured an Intel Core i7-4600MQ CPU, an NVIDIA GT 730M GPU, and 8GB DDR3 RAM.

The detailed running times for each experiment are provided in Table 3. Since MFAPC-CFDL uses a recursive formula to update the control law, it has a shorter computation burden, as shown in Table 3. However, the other methods require solving the optimization problem at each sampling time, increasing the computational load.

It can be observed from Figure 1 that the WKPC strategy exhibits a more aggressive response compared to other methods, resulting in maximum overshoot and minimum rise-time. On the other hand, MFAPC acts smoothly based on Figure 3 and has a minimum rate of change in control signals. Furthermore, MFAPC has approximately zero steady error, but the other two methods have steady-state error.

*Table 2. MFAPC-CFDL parameters*

| $N$ | $\lambda$ | $\rho$ | $\mu$ | $\eta$ |
|---|---|---|---|---|
| 0.5[s] | 0.37 | 1 | 1 | 1 |
| $\varepsilon$ | $\delta$ | $\hat{\phi}_0$ | $\hat{\theta}_0$ | $n_m$ |
| $10^{-5}$ | 0.5 | 0.1 | $0.175 \times [1,1,1,1]$ | 4 |

*Table 3. Numerical example results*

| | **MFAPC-CFDL** | **DeePC** | **WKPC** |
|---|---|---|---|
| Absolute integral error (°) | 730.4181 | 606.1664 | 456.3552 |
| Minimum absolute error (°) | $7.7326 \times 10^{-4}$ | 0.2464 | 0.1662 |
| Maximum absolute input ($\boldsymbol{Nm}$) | 3.3593 | 3.4353 | 3.5 |
| Optimization time ($s$) | $6.55 \times 10^{-5}$ | 0.0111 | 0.0117 |

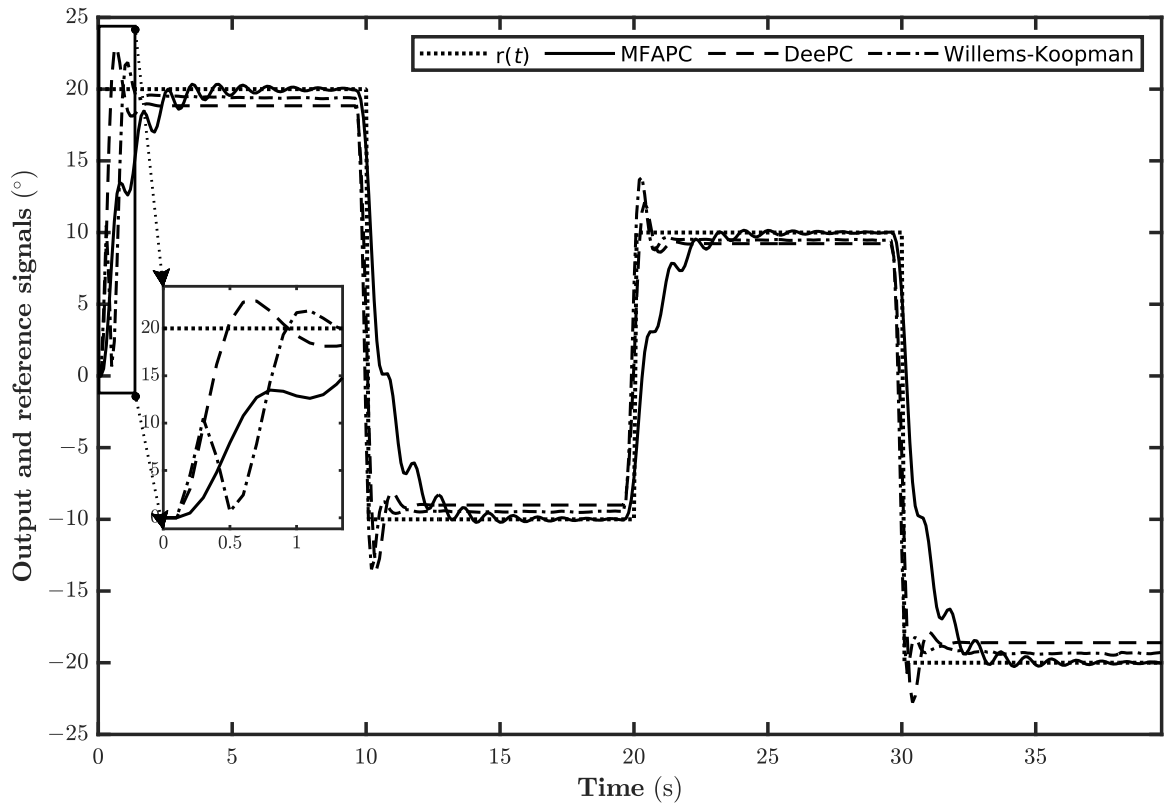

*Figure 1. Closed-loop response Comparison*

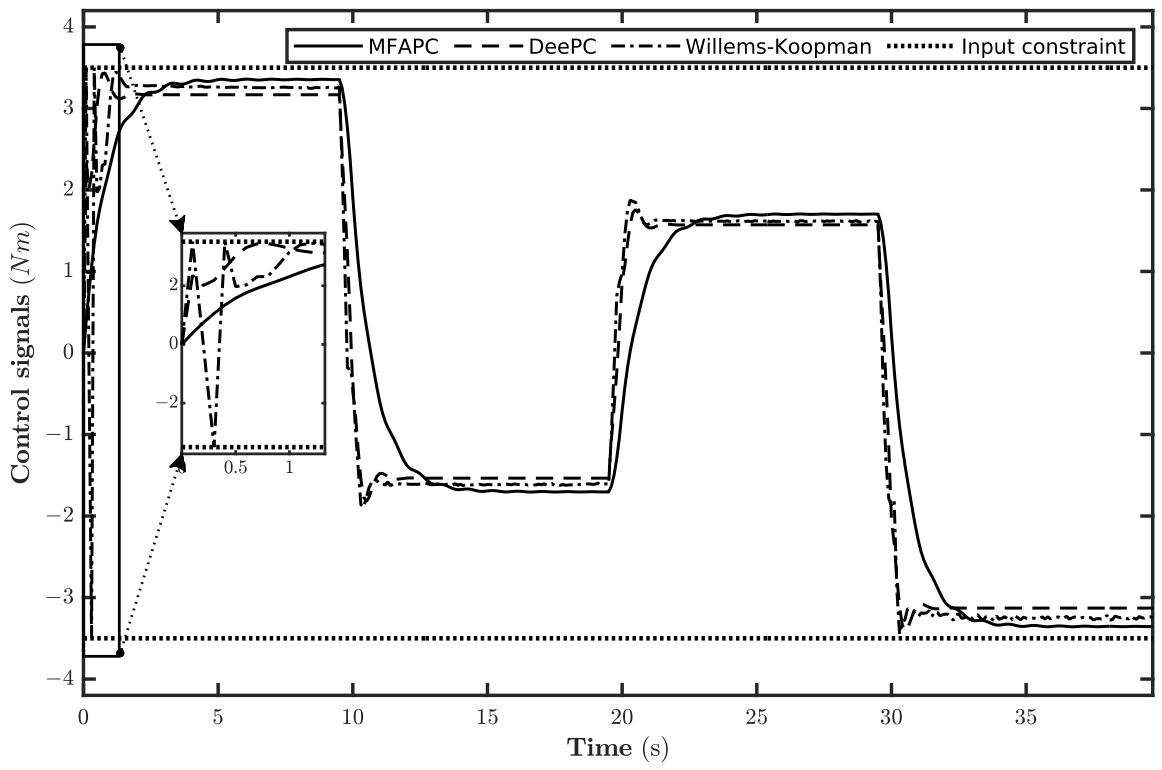

*Figure 2. Control signals comparison*

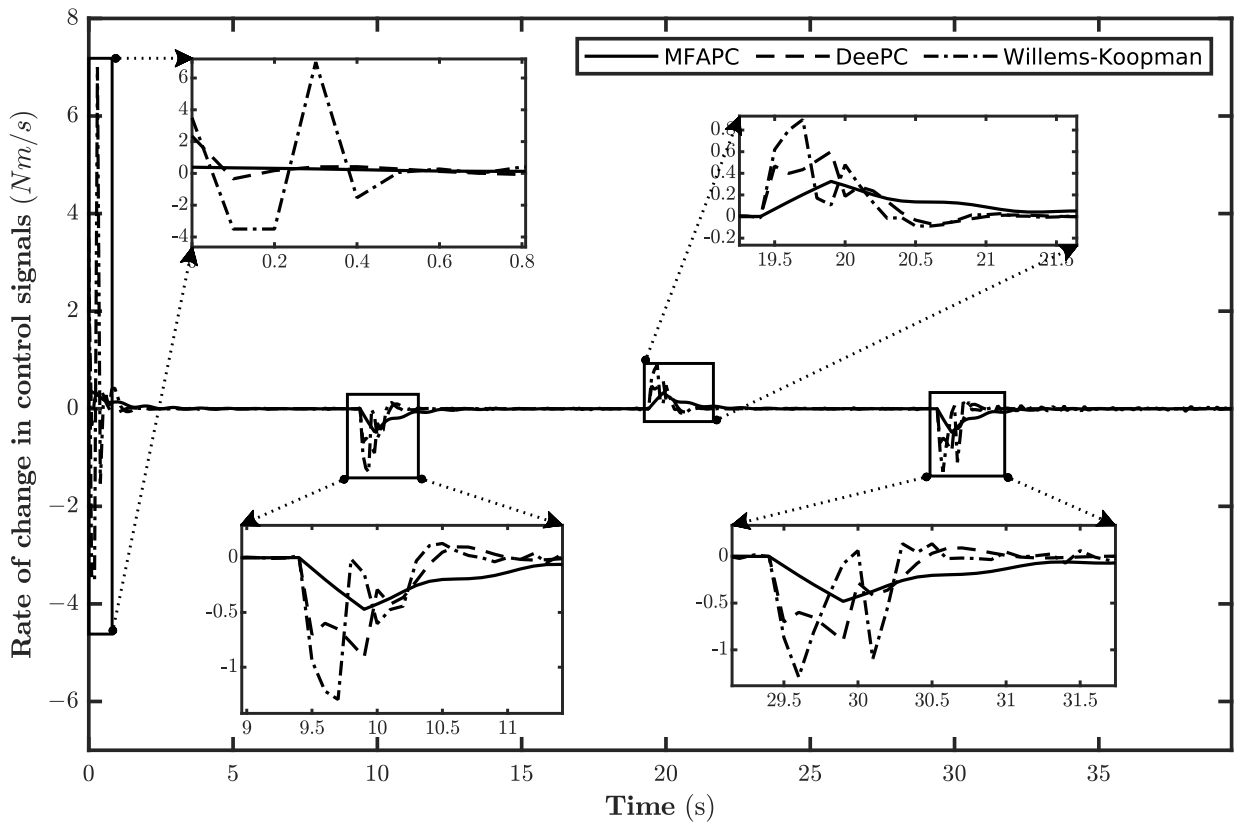

*Figure 3. Comparison of the rate of change in control signals*

For the DeePC algorithm, empirical findings reveal that $\lambda_g$ and $\lambda_\sigma$ must fall within an appropriate range, where their amplitude insignificantly influences control performance. In contrast, the intricate interaction between $S$, $R$, and $Q$, and their ratios relative to $\lambda_g$ and $\lambda_\sigma$, plays a pivotal role in determining control performance, necessitating precise fine-tuning to achieve desired outcomes.

Similarly, in the case of the WKPC algorithm, the number of lifted states $n_p$ and $\lambda_g$ should primarily be maintained within their appropriate range, as within this range, they are observed to have a negligible effect on control performance. However, the ratios of $R$ and $Q$ to each other and their relationship to $\lambda_g$ significantly impact control performance and necessitate meticulous fine-tuning to achieve optimal results.

It is important to emphasize that all results and conclusions in this part are derived from empirical studies conducted through simulations.

The simulation codes used in this study can be found at "https://github.com/Sohrab-rz/Comparative-Analysis-of-Data-Driven-Predictive-Control-Strategies".

## V. Conclusion

This paper comprehensively examines data-driven predictive control strategies, examining their theories, fundamental assumptions, and applications. DeePC, WKPC, and MFAPC-CFDL's operating principles have been explained. Through rigorous simulation studies, this paper has highlighted these strategies' relative strengths and limitations. DeePC and WKPC exhibit similar behaviors, with WKPC having a more complicated structure, which results in improved performance over DeePC. Still, some assumptions must be considered for these methods to be compatible with time-varying systems. On the other hand, MFAPC, with its simple structure, offers adaptability to system changes but needs to improve in performance compared to the other two methods. Moreover, MFAPC also requires a lower PE order for the input signal, which is a significant advantage.

## Acknowledgment

The members of the Advanced Control Systems Laboratory (ACSL) at K. N. Toosi University of Technology have been supportive throughout this research.

## References

[1] E. F. Camacho, C. Bordons, E. F. Camacho, and C. Bordons, Model predictive controllers. Springer, 2007.

[2] Zhongsheng Hou and Shangtai Jin. *Model free adaptive control: theory and applications*. CRC press, 2013.

[3] Zhong-Sheng Hou and Zhuo Wang. From model-based control to data-driven control: Survey, classification and perspective. *Information Sciences*, 235:3–35, 2013.

[4] Wang Jianhong, Ricardo A Ramirez-Mendoza, and Ruben Morales-Menendez. *Data Driven Strategies: Theory and Applications*. CRC Press, 2023.

[5] Anjukan Kathirgamanathan, Mattia De Rosa, Eleni Mangina, and Donal P Finn. Data-driven predictive control for unlocking building energy flexibility: A review. *Renewable and Sustainable Energy Reviews*, 135:110120, 2021.

[6] Ali Khaki-Sedigh. *An Introduction to Data-Driven Control Systems*. Wiley-IEEE Press, 2024.

[7] Jan C Willems, Paolo Rapisarda, Ivan Markovsky, and Bart LM De Moor. A note on persistency of excitation. *Systems & Control Letters*, 54(4):325–329, 2005.

[8] J. Coulson, J. Lygeros, and F. Dörfler, "Data-enabled predictive control: In the shallows of the DeePC," in 2019 18th European Control Conference (ECC), 2019: IEEE, pp. 307-312.

[9] Julian Berberich, Johannes Köhler, Matthias A. Müller, and Frank Allgöwer, "Linear tracking MPC for nonlinear systems—Part II: The data-driven case," IEEE Transactions on Automatic Control, vol. 67, no. 9, pp. 4406-4421, 2022.

[10] Igor Mezić and Andrzej Banaszuk. Comparison of systems with complex behavior. *Physica D: Nonlinear Phenomena*, 197(1-2):101–133, 2004.

[11] Igor Mezić. Spectral properties of dynamical systems, model reduction and decompositions. *Nonlinear Dynamics*, 41:309–325, 2005.

[12] Ivan Markovsky and Florian Dörfler. Behavioral systems theory in data-driven analysis, signal processing, and control. *Annual Reviews in Control*, 52:42–64, 2021.

[13] Steven L Brunton. Notes on koopman operator theory. *Universita ̈t von Washington, Department of Mechanical Engineering, Zugriff*, 30, 2019.

[14] Julian Berberich, Johannes Köhler, Matthias A Müller, and Frank Allgöwer. Data-driven model predictive control with stability and robust- ness guarantees. *IEEE Transactions on Automatic Control*, 66(4):1702– 1717, 2020.

[15] Tahereh Gholaminejad and Ali Khaki-Sedigh. Stable data-driven koop- man predictive control: Concentrated solar collector field case study. *IET Control Theory & Applications*, 17(9):1116–1131, 2023.

[16] Yuan Guo, Zhongsheng Hou, Shida Liu, and Shangtai Jin. Data-

driven model-free adaptive predictive control for a class of mimo nonlinear discrete-time systems with stability analysis. *IEEE Access*, 7:102852– 102866, 2019.

[17] Feilong Zhang. Data-driven model-free adaptive predictive control and its stability analysis. *arXiv preprint arXiv:1910.08321*, 2019.

[18] Shida Liu, Zhongsheng Hou, and Jian Zheng. Attitude adjustment of quadrotor aircraft platform via a data-driven model free adaptive control cascaded with intelligent PID. In *2016 Chinese Control and Decision Conference (CCDC)*, pages 4971–4976. IEEE, 2016.